# WDM for Multi-user Indoor VLC Systems with SCM

Safwan Hafeedh Younus[1*], Aubida A. Al-Hameed[1,] Ahmed Taha Hussein[1], Mohammed T. Alresheedi[2] and Jaafar M. H. Elmirghani[1]

[1]School of Electronic and Electrical Engineering, University of Leeds, LS2 9JT, United Kingdom
[2]Department of Electrical Engineering, King Saud University, Riyadh, Kingdom of Saudi Arabia
[*]elshy@leeds.ac.uk

**Abstract:** A system that employs wavelength division multiplexing (WDM) in conjunction with subcarrier multiplexing (SCM) tones is proposed to realize high data rate multi-user indoor visible light communication (VLC). The SCM tones, which are unmodulated signals, are used to identify each light unit, to find the optimum light unit for each user and to calculate the level of the co-channel interference (CCI). WDM is utilized to attain a high data rate for each user. In this paper, multicolour (four colours) laser diodes (LDs) are utilized as sources of lighting and data communication. One of the WDM colours is used to convey the SCM tones at the beginning of the connection to set up the connection among receivers and light units (to find the optimum light unit for each user). To evaluate the performance of our VLC system, we propose two types of receivers: an array of non-imaging receivers (NI-R) and an array of non-imaging angle diversity receivers (NI-ADR). In this paper, we consider the effects of diffuse reflections, CCI and mobility on the system performance.

## 1. Introduction

Managing multi-user communication in VLC systems is challenging due to use of many transmitters in VLC systems to obtain an acceptable illumination level that meets the lighting demands. The overlap in coverage area between the multiple transmitters is therefore very high [1]. Furthermore, the communication area (size of the cell) in VLC systems changes based on the illumination level provided, which means that the multiple VLC transmitters in the room are usually treated as one source of information/transmitter [2]. In addition, intensity modulation and direct detection (IM/DD) are the most preferred modulation technique in VLC systems. Therefore, transmitting many signals through overlapping multiple transmitters increases interference from undesired transmitters [1].

Many approaches have been proposed to provide multi-user communication in VLC systems. One of these is to use multiple input single output (MISO) and multiple input multiple output (MIMO) techniques with zero-forcing (ZF) precoding schemes to reduce the effects of co-channel interference (CCI) between transmitters [3]-[5]. Other approached utilize multiple access dimensions to support multi-user communication in VLC systems. The methods investigated include code division multiple access (CDMA) [6], non-orthogonal multiple access (NOMA) [7], [8] and expurgated pulse position modulation (EPPM) [8]. A range of resource allocation methods were also developed for multi-user communication in VLC systems [9]-[11].

In this paper, WDM in conjunction with subcarrier multiplexing (SCM) tones is used to realise a high data rate multi-user indoor VLC system where the spatial domain and the relatively high directivity of optical signals are also exploited by assigning a light unit to each user. SCM tones studied in indoor optical wireless communication systems for a range of applications [12]-[14]. In [12], the SCM tones were utilized to find the crosstalk level between WDM channels while in [13], the SCM tones were utilized to design an indoor VLC positioning system. The work [14] introduced SCM tones for parallel data transmission in an indoor VLC system where the SCM tones were utilized to match each light unit with the pixel(s) of an imaging receiver as well as to calculate the CCI level. We use SCM tones here to recognise each light unit, to find the best light unit for each user and to calculate the level of the CCI. RYGB laser diodes (LDs) are utilized as sources of illumination and data communication. At the beginning of the connection, one colour of the RYGB LDs light unit (green colour here) sends the SCM tones to set up the connection between the light units and the users. After identifying the optimum light unit for each user, data is transmitted in parallel through RYGB LDs. Two types of receivers are proposed in this work: an array of non-imaging receivers (NI-R) and an array of non-imaging angle diversity receivers (NI-ADR). The performance of our proposed system is investigated in an empty room taking into account the effects of diffuse reflections, CCI and mobility using simple on-off-keying (OOK) modulation.

## 2. VLC system model

The empty room shown in Fig. 1 (a) is used to investigate the performance of our proposed system. This room has an area equal to 4 m × 8 m and height of 3 m. Reflecting surfaces (walls, ceiling and floor) of the room are presumed to be Lambertian reflectors [15]. Thus, the walls, the ceiling and the floor of the room were modeled as Lambertian reflectors with reflection coefficients of 0.8 for the walls and ceiling and 0.3 for the floor. In this paper, we consider reflections up to second order [16], [17]. To model the reflections from the reflecting surfaces of the room, a ray tracing algorithm is utilized. Thus, the room is split into small square elements (surface elements) that have an area $d_A$ and a reflection coefficient of $\rho$. These surface elements are treated as secondary emitters that reflected signals in the form of a Lambertain pattern with $n = 1$, where $n$ is the Lambertain emission order. The area of the small elements was chosen to be 5 cm × 5 cm for the first order reflections and 20 cm × 20 cm for the second order reflections [18].

To obtain high modulation bandwidth, we use RYGB LDs instead of LEDs. Experimental results have revealed that multicolour LDs can be utilized for illumination without any hazards on the human eye [19], [20]. Twelve light units have been installed on the ceiling and each light unit has 6 RYGB LDs (2 × 3) to achieve sufficient illumination level in our proposed environment, which is not less than 300 lx [21]. Due to the use a diffuser, the output of the RYGB LDs has a Lambertian radiation pattern. Calculation of the illumination level can be found in [22]-[24]. Fig. 1 (b) shows the configuration of the RYGB LDs [25]. The green colour sends the SCM tones to set up the connection and subsequently, the data is transmitted in parallel through the four colours of the RYGB LDs.

Two types of receivers are proposed in this work: the NI-R and the NI-ADR receivers as shown in Fig. 2 (a) and (b). In the NI-R, an array (2 × 2) of photodetectors is used. Each photodetector has an area equal to 6.25 mm$^2$ and a FOV of 40º. The area of each photodetector in the NI-R is selected to enable each photodetector to work at a high data rate while gathering a significant optical power. Calculations of the bandwidth of the photodetector can be found in [26]. The FOV of each photodetector is chosen to ensure that the NI-R sees at least one light unit at any location of the optical receiver on the communication floor. As shown in Fig. 2 (a), each photodetector is covered with a different optical filter. Hence, each photodetector responds to a specific wavelength. Due to the use of four colours LDs as sources of illumination and data communication, four colours optical filters (red, yellow, green and blue) are used as shown in Fig. 2 (a). The green colour is used to send the SCM tones at the beginning of the connection, although any other colour can be used. Thus, at the beginning of the connection, the output of the green photodetector enters to the SCM tone identification system to set up the connection. When each user is assigned to its optimum light unit, the output of the green colour enters to the parallel to serial (P/S) converter, as this colour is then used to send a data stream as shown in Fig 2 (a).

To reduce the effect of CCI and the multipath reflections on the performance of our proposed system, the NI-ADR is introduced. This NI-ADR has seven faces (1 − 7) and each face consists of an array of four photodetectors (2 × 2) with an area of 4 mm$^2$ for each photodetector as shown in Fig. 2 (b). Each photodetector is covered by a different optical filter and consequently, each photodetector senses a different wavelength. To determine the orientation of each branch in the NI-ADR, two angles are used: azimuth ($A_z$) and elevation ($El$). The $El$ angle of the first face is selected to be equal to 90º, while the other six branches are given an $El$ of 50º. The $A_z$ angle is defined as the orientation of the branch's angle. The $A_z$ angles of the seven faces are fixed at 0º, 0º, 60º, 120º, 180º, 240º and 300º. In addition, the FOV of each photodetector is selected to be 20º. It should be noted that the parameters of the NI-ADR ($A_z$, $El$ and FOV) are chosen to ensure that the NI-ADR is able to see at least one transmitter at any location of the optical receiver on the communication floor of the room following an optimisation similar to that in [27]. Each photodetector in each face of the NI-ADR amplifies the received signal separately; therefore, many possible diversity schemes can be used: select the best (SB) scheme, equal gain combining (EGC) scheme and maximum ratio combining (MRC) scheme. We used SB for the photodetectors that covered by the same colour filters as shown in Fig. 2 (b).

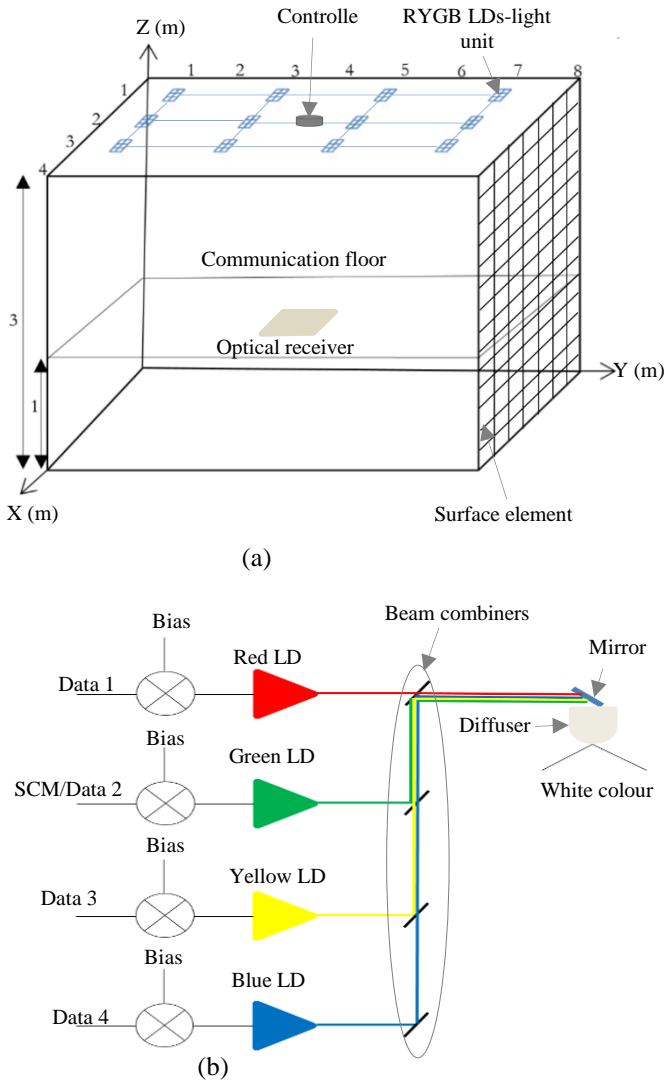

Fig. 1: (a) proposed room configuration and (b) RYGB LD configuration.

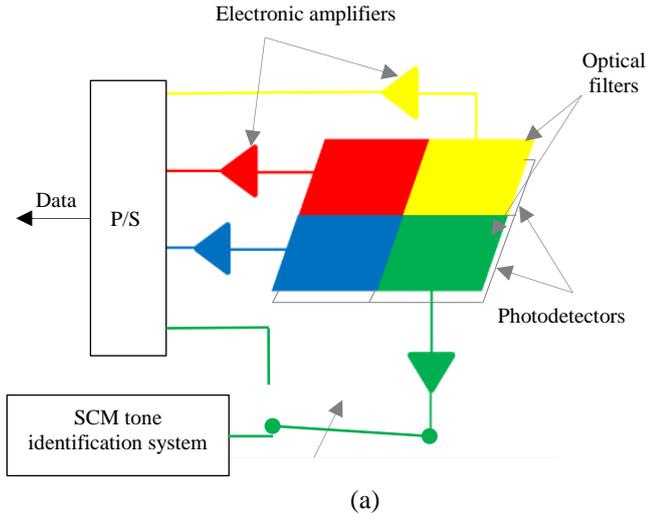

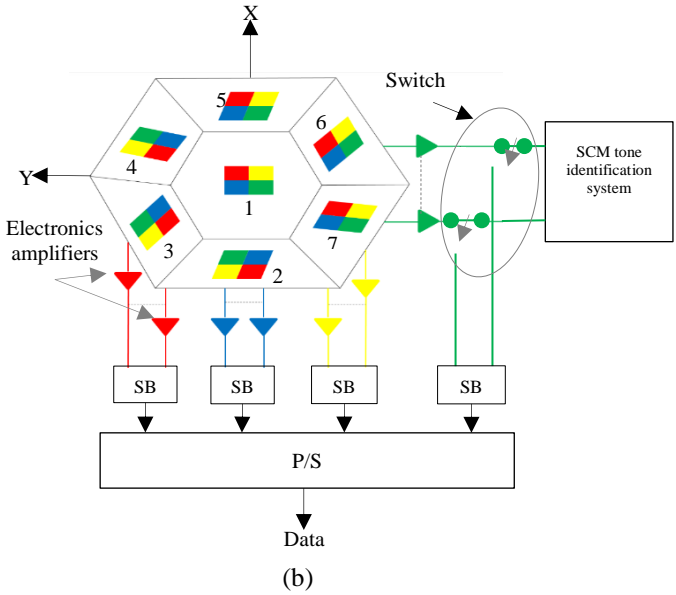

*Fig. 2: Receivers structure: (a) NI-R configuration and (b) NI-ADR configuration.*

Table 1 shows the parameters that were used in the simulation.

Table 1: Simulation parameters.

| Parameters | Configurations | |
|---|---|---|
| **Room** | | |
| Length | 8 m | |
| Width | 4 m | |
| Height | 3 m | |
| $\rho$-xz Wall | 0.8 | |
| $\rho$-yz Wall | 0.8 | |
| $\rho$-xz op. Wall | 0.8 | |
| $\rho$-yz op. Wall | 0.8 | |
| $\rho$-Floor | 0.3 | |
| Bounces | 1 | 2 |
| Number of elements | 32000 | 2000 |
| $d_A$ | 5 cm × 5 cm | 20 cm × 20 cm |
| Lambertian emission order ($n$) | 1 | |
| Sime-angle at half power | 60º | |
| **Transmitters (RYGB LDs-light units)** | | |
| Number of transmitters | 12 | |
| Transmitters/light units locations($x$, $y$, $z$) m | (1, 1, 3), (1, 3, 3), (1, 5, 3),(1, 7, 3), (2, 1, 3), (2, 3, 3), (2, 5, 3), (2, 7, 3), (3, 1, 3), (3, 3, 3), (3, 5, 3), (3, 7, 3) | |
| Number of RYGB LDs per unit | 6 (2 × 3) | |
| RYGB LDs interval | 0.02 m | |
| Optical power of R LD | 0.8 W | |
| Optical power of Y LD | 0.5 W | |
| Optical power of G LD | 0.3 W | |
| Optical power of B LD | 0.3 W | |
| Total centre luminous intensity | 162 cd | |
| Lambertian emission order ($n$) | 0.65 | |
| Semi-angle at half power | 70º | |
| **NI-R** | | |
| Number of photodetectors | 4 | |
| FOV of each photodetector | 40º | |
| Elevation of each photodetector | 90º | |
| Azimuth of each photodetector | 0º | |
| Photodetector's area | 6.25 mm² | |
| Photodetector's responsivity (red) | 0.4 | |
| Photodetector's responsivity (yellow) | 0.35 | |
| Photodetector's responsivity (green) | 0.3 | |
| Photodetector's responsivity (blue) | 0.2 | |
| Receiver's bandwidth | 0.5 GHz | |
| **NI-ADR** | | |
| Number of branches | 7 | |
| Number of photodetectors/branch | 4 | |
| FOV of each photodetector | 20º | |
| Elevation of each branch | 90º, 50º, 50º, 50º, 50º, 50º, 50º | |
| Azimuth of each branch | 0º, 0º, 60º, 120º, 180º, 240º 300º | |
| Photodetector's area | 4 mm² | |
| Receiver's bandwidth | 0.75 Hz | |

## 3. System Description

In the multi-user scenario, interference happens when multiple signals from the interfering light units land on the optical receiver. The interference is occurred due to either LOS components and/or non-line-of-sight (NLOS) components. The optical receiver of each user is connected with all light units through the SCM tones. Thus, the output of the photodetectors that covered by the green filters enter to twelve bandpass filters (BPFs) as shown in Fig. 3. This BPFs have center frequencies equal to the frequencies of the SCM tones.

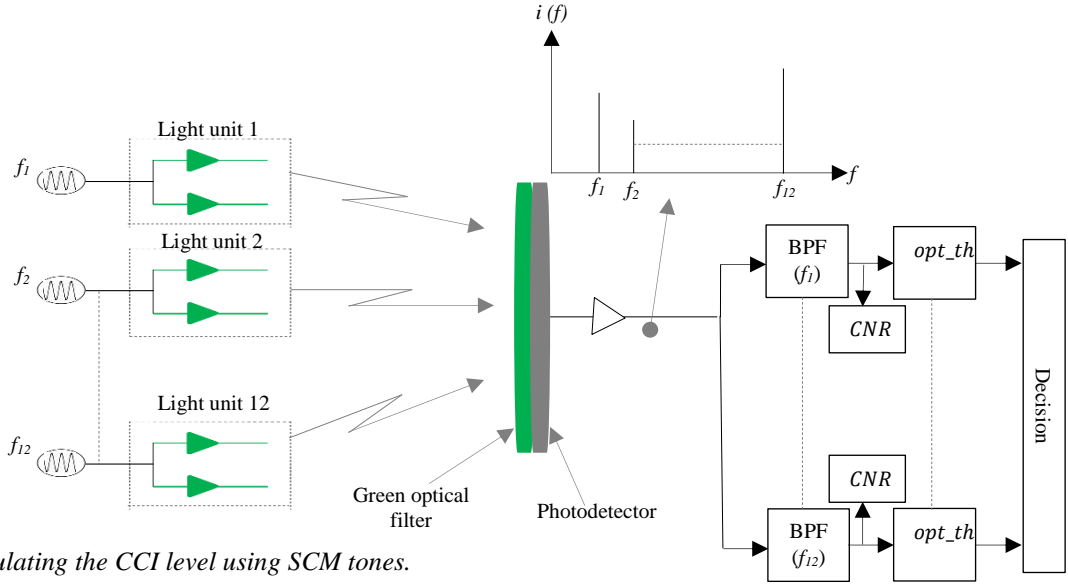

*Fig. 3: Calculating the CCI level using SCM tones.*

The indoor channel frequency response of VLC systems has low pass characteristic in the electrical domain [28]. Thus, the range of the frequencies assigned to the SCM tones should be selected near to the DC where the channel has low attenuation. We obtained the 3-dB channel bandwidth of the NI-R and the NI-ADR when the mobile user moved in steps of 1 m along the $y$-axis and at $x = 0.5$ m and $x = 1.5$ m as shown in Table 2. We considered locations at $x = 0.5$ m and at $x = 1.5$ and along the $y$-axis as the user moves near to the wall (at $x = 0.5$ m) and moves between light units not underneath them (at $x = 1.5$ m) where inter-symbol interference (ISI) is high. It can be seen that the minimum 3-dB channel bandwidth is associated with the NI-R, and is 1.23 GHz. Therefore, the frequency range selected for the 12 SCM tones was 500 MHz to 1160 MHz with a guard of 60 MHz. The bandwidth of each BPF was selected equal to 4 MHz, which reduces the noise that seen by each SCM tone, and allows for SCM oscillator drift and BPF tolerances [12].

Table 2: The 3-dB channel bandwidth of the NI-R and the NI-ADR when the mobile user moved along the $y$-axis and at $x = 0.5$ m and $x = 1.5$ m.

| Y(m) | 3-dB Channel bandwidth (GHz) | | | |
|---|---|---|---|---|
| | Receiver at x = 0.5 m | | Receiver at x = 1.5 m | |
| | NI-R | NI-ADR | NI-R | NI-ADR |
| 0.5 | 1.23 | 3.37 | 1.56 | 4.63 |
| 1.5 | 1.64 | 4.66 | 1.8 | 4.7 |
| 2.5 | 1.93 | 5.1 | 2.4 | 5.3 |
| 3.5 | 2.05 | 5.2 | 2.5 | 5.6 |
| 4.5 | 2.05 | 5.2 | 2.5 | 5.6 |
| 5.5 | 1.93 | 5.1 | 2.4 | 5.3 |
| 6.5 | 1.64 | 4.66 | 1.8 | 4.7 |
| 7.5 | 1.23 | 3.37 | 1.56 | 4.63 |

### 3.1 Transmitters Allocation

A controller connected to all light units (see Fig. 1 (a)) is used to manage the light units and the users. Each user is assigned one light unit for data transmission and the other light units (the light units assigned to other users) are considered as interfering light units. Thus, based on the channel conditions between the transmitters and the receivers, each user is allocated the light unit that has a good channel (high optical received power and low level of CCI). To find the optimum light unit that gives a reliable connection to each user, the SCM tone identification system at each receiver is used to calculate the received power of each SCM tone.

The SCM tones are separated in frequency; thus, there is no interference between the SCM tones. Therefore, the output of each BPF (see Fig. 3) is a SCM tone plus noise. Hence, to match each user with its closest light unit (the light unit that offers the highest optical power), an optimum threshold should be set. When the NI-Rs are used, in some user locations, more than one light units (up to four light units) have LOS components with the users and this is attributed to the wide FOV of the NI-R and the distribution of the light units. In this case, the ability of the SCM tone identification system to match each user with its closest light unit should be evaluated.

We defined three cases based on the location of a user. In case one, a user has LOS components with one light unit and has just NLOS components with the other light units. In case two, a user has LOS components with two light units and has just NLOS components with the other light units. In case three, a user has LOS components with three or four light units and has just NLOS components with the other light units. In addition, we defined the probability of not allocating the closest light unit to each user as the probability of a wrong decision. Thus, in case one, the probability of wrong decision is the probability of a user being assigned a light unit that has no LOS components with it. While in case two, the probability of the wrong decision is the probability of assigning the second closest light unit to a user, in case three, the probability of a wrong decision is the probability of assigning the second, the third or the fourth nearest light unit to a user. Here, we will evaluate the probability of wrong decision for case two only. This is due to the fact that in case two, the probability of allocating the second nearest light uint to a user is higher than the probability of assigning a light unit that has no LOS components with a user (case one) and higher than the probability of assigning the third or the fourth closest light unit to a user (case three). This is because we consider the received power of each light unit to decide which light unit offers the highest power.

In case two, the output of each BPF (see Fig. 3) is the SCM tone transmitted from the closest light unit (desired SCM tone) to a user plus noise, the SCM tone transmitted from the second nearest light unit (undesired SCM tone) to a user plus noise or the SCM tone that reaches the user through only NLOS components plus noise, which can be ignored (case one). An optimum threshold should be obtained to decide the output of each the BPF.

To classify the electrical current ($z$) output of each BPF, we hypothesize two states [14]. Under hypothesis 1 ($H1$), $z$ is the undesired SCM tone plus noise ($b + n$), while under hypothesis 2 ($H2$), $z$ is the desired SCM tone plus noise ($b + n$).

The total noise ($n$) that is seen by each SCM tone is white Gaussian zero mean, with total standard deviation $\sigma_t$ that can be written as [29]:

$$\sigma_t = \sqrt{\sigma_{bn}^2 + \sigma_s^2 + \sigma_{pr}^2} \qquad (1)$$

where $\sigma_{bn}$ is the background shot noise component, $\sigma_s$ is the shot noise component associated with the received SCM tone and $\sigma_{pr}$ is the preamplifier noise component. Calculation of the $\sigma_{bn}$ and $\sigma_s$ can be found in [15], [30]. In this paper, we asummed the p-i-n FET receiver design in [31], which has an input noise current equal to 4.5 pA/$\sqrt{Hz}$.

The values of the desired SCM tone electrical current ($a$) and the undesired SCM tone electrical current ($b$) depend on the distance between the optical receiver and the light units. Therefore, $a$ and $b$ are random variables and their distributions were obtained through simulation of the indoor channel. The simulations considered 1000 random positions of the receiver on the communication floor to determine the distribution of $a$ and $b$. At each position of the receiver, the values of $a$ and $b$ were calculated. Fig. 4 (a) and Fig. 4 (b) show the histogram and the curve fitting of the $a$ and $b$ parameters respectively.

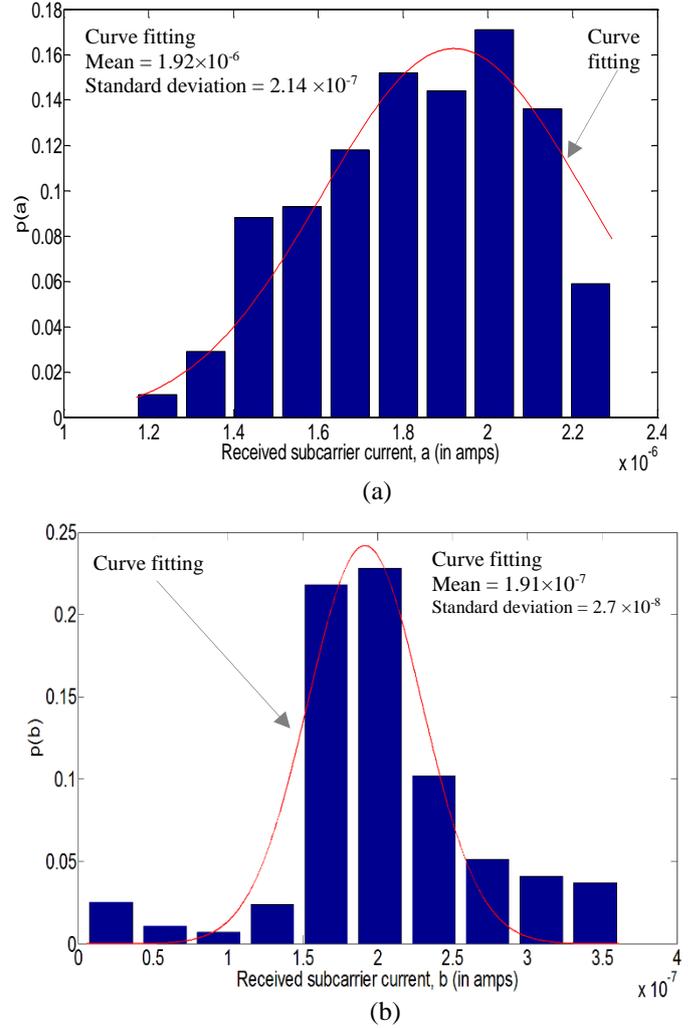

Fig. 4: Histogram and curve fitting of the (a) desired SCM tone received electrical current and (b) undesired SCM tone received electrical current.

From the curve fitting, the normalized probability density functions (pdfs) of $a$ ($p(a)$) and $b$ ($p(b)$) are written as:

$$p(a) = \frac{1}{\sqrt{2\pi}\,\sigma_{ds}} e^{-\left(\frac{a-m_{ds}}{\sqrt{2}\sigma_{ds}}\right)^2} \qquad (2)$$

and

$$p(b) = \frac{1}{\sqrt{2\pi}\,\sigma_{us}} e^{-\left(\frac{b-m_{us}}{\sqrt{2}\sigma_{us}}\right)^2} \qquad (3)$$

where we presumed a Gaussian distribution for the received power which is reasonable given the multiple locations (giving rise to multiple angles and distances), the multiple reflecting surfaces and the observed results. Here, $\sigma_{ds}$ and $m_{ds}$ are the standard deviation and the mean value of $a$, respectively and $\sigma_{us}$ and $m_{us}$ are the standard deviation and the mean value of $b$, respectively.

The pdfs of $z$ given $H1$ and $H2$ can be written as follows: under $H1$, $z$ is the convolution of the undesired SCM tone pdf and the noise pdf:

$$fz(z|H1) = p(b) \otimes p(n) \qquad (4)$$

Solving equation (4), $fz(z|H1)$ can be written as:

$$fz(z|H1) = \frac{1}{\sqrt{2\pi(\sigma_{us}^2 + \sigma_t^2)}} e^{-\left(\frac{z-m_{us}}{\sqrt{2(\sigma_{us}^2+\sigma_t^2)}}\right)^2} \quad (5)$$

Under $H2$, $z$ is the convolution of the desired SCM tone pdf and the noise pdf, which is given as:

$$fz(z|H2) = \frac{1}{\sqrt{2\pi(\sigma_{ds}^2 + \sigma_t^2)}} e^{-\left(\frac{z-m_{ds}}{\sqrt{2(\sigma_{ds}^2+\sigma_t^2)}}\right)^2} \quad (6)$$

Applying likelihood ratio to equations (5) and (6), we get:

$$\frac{fz(z|H2)}{fz(z|H1)} \underset{H1}{\overset{H2}{\gtreqless}} 1$$

$$= \frac{\frac{1}{\sqrt{2\pi(\sigma_{ds}^2 + \sigma_t^2)}} e^{-\left(\frac{z-m_{ds}}{\sqrt{2(\sigma_{ds}^2+\sigma_t^2)}}\right)^2}}{\frac{1}{\sqrt{2\pi(\sigma_{us}^2 + \sigma_t^2)}} e^{-\left(\frac{z-m_{us}}{\sqrt{2(\sigma_{us}^2+\sigma_t^2)}}\right)^2}} \underset{H1}{\overset{H2}{\gtreqless}} 1 \quad (7)$$

Solving equation (7), we get:

$$z \underset{H1}{\overset{H2}{\gtreqless}} \frac{1}{\sigma_{ds}^2 - \sigma_{us}^2} \left( m_{us}(\sigma_{ds}^2 + \sigma_t^2) - m_{ds}(\sigma_{us}^2 + \sigma_t^2) + \sqrt{\begin{pmatrix} \sigma_{ds}^2\sigma_{us}^4 + \sigma_t^4\sigma_{us}^2 + \sigma_t^2\sigma_{us}^4 \\ -\sigma_{ds}^4\sigma_t^2 - \sigma_{ds}^4\sigma_{us}^2 - \sigma_{ds}^2\sigma_t^4 \end{pmatrix} \ln\left(\frac{\sigma_{us}^2+\sigma_t^2}{\sigma_{ds}^2+\sigma_t^2}\right) + (m_{ds} - m_{us})^2(\sigma_t^2\sigma_{ds}^2 + \sigma_{ds}^2\sigma_{us}^2 + \sigma_t^4 + \sigma_t^2\sigma_{us}^2)} \right) =$$

$$z \underset{H1}{\overset{H2}{\gtreqless}} opt\_th \quad (8)$$

If $m_{ds}$ is very large compared with $\sigma_{ds}$, $\sigma_{us}$ and $m_{us}$, the optimum threshold ($opt\_th$) ~ $\frac{m_{ds}}{2}$ as expected. The probability of correct detection of the desired SCM tone, which is the probability of a correct decision on the desired SCM tone ($Pcds$) is:

$$Pcds = \int_{opt\_th}^{\infty} fz(z|H2)dz$$

$$= \int_{opt\_th}^{\infty} \frac{1}{\sqrt{2\pi(\sigma_{ds}^2 + \sigma_t^2)}} e^{-\left(\frac{z-m_{ds}}{\sqrt{2(\sigma_{ds}^2+\sigma_t^2)}}\right)^2} dz \quad (9)$$

The probability of correct detection of the undesired SCM tone, which is the probability of a false alarm in the undesired SCM tone ($Pfus$) (i.e. the undesired SCM tone is considered as the desired SCM tone) is:

$$Pfus = \int_{opt\_th}^{\infty} fz(z|H1)dz$$

$$= \frac{1}{\sqrt{2\pi(\sigma_{us}^2 + \sigma_t^2)}} e^{-\left(\frac{z-m_{us}}{\sqrt{2(\sigma_{us}^2+\sigma_t^2)}}\right)^2} dz \quad (10)$$

Consequently, the probability of not allocating the undesired SCM tone to a user, which is the probability of correct decision in the undesired SCM tone ($Pcus$) is:

$$Pcus = 1 - Pfus \quad (11)$$

Thus, the overall probability of the SCM identification system makes a correct dissection ($Pcd$) is:

$$Pcd = Pcds\,(Pcus)^{M-1} \quad (12)$$

where $M$ is the number of the light units. Consequently, the probability of making wrong decision in the SCM identification system is $Pwd$ is $1 - Pcd$. In our system, and for the given set of parameters in this paper, $Pwd$ is $1.2 \times 10^{-12}$, which shows that the SCM tone identification system has the ability to match each user to its nearest light unit with high accuracy. Due to the distribution of the light units on the ceiling of the room, in some user locations, two or four light units may have the same distances to a user. In this case, any one of these light units can be considered as the nearest one to a user.

When NI-ADRs are used instead of NI-Rs, each photodetector covered by the green filter in each face of the NI-ADR is treated separately to find the nearest light unit to each user. However, the FOV of each photodetector is 20°, which means no more than one light unit (the light unit that has LOS components with the photodetector) can be seen by each face. This is similar to case one in the NI-R, which means that the probability of wrong decision is lower than $10^{-12}$ for the NI-ADR.

By calculating the carrier to noise ($CNR$) ratio of each SCM tone at each user, through an infrared (IR) signal, the optical receiver of each user informs the controller the $CNR$ associated with each SCM tone. Each user is allocated one light unit. For each user, the controller sorts the light units in descending order (each user has a different descending order of the light units starting with the light unit that has the highest ($CNR$) and ending with the light unit that has the lowest ($CNR$)). The controller allocates one light unit, the light unit that has the highest $CNR$, to each user to transmit the data. It should be noted that we used the design of the IR uplink presented in [32]. In addition, to prevent interference in the uplink, each user is given a time slot to send the feedback information to the controller. The electrical power of a SCM tone ($S$) at a user is given as:

$$S_{m,n} = \left(\frac{R_g\,Pr_{m,n}}{2}\right)^2 \quad (13)$$

here $R_g$ is the photodetector's responsivity for the green spectrum and $Pr_{m,n}$ is optical (green) power received at $n^{th}$

user due to $m^{th}$ transmitter. Consequently, $CNR$ of any SCM tone at any user is given as:

$$CNR_{n,m} = \frac{(R_g\, Pr_{m,n})^2}{2\,\sigma_t^2} \quad (14)$$

Due to the fact that the SCM tones are unmodulated tones and have unique frequencies, there is no interference between these SCM tones at any optical receiver. Therefore, the CCI level is defined as the total received power at $n^{th}$ receiver, except for the received power of the desired SCM tone. For example, if the desired tone of $m^{th}$ transmitter is $f_m$, the level of the CCI at the green photodetector of $n^{th}$ user ($I_{n\_green}$) due to the other SCM tones are given as:

$$I_{n\_green} = \sum_{\substack{k=1 \\ k \ne m}}^{Ma} (\frac{R_g\, Pr_{n,k}}{2})^2, \quad n \in [1, 2 \ldots N], \quad (15)$$
$$m \in [1, 2 \ldots M]$$

where $Ma$ is the number of active light units (the light units assigned to users to send data) and $N$ is the number of receivers (in our system, $N = Ma$).

The closest light unit to a user offers a higher optical power than the other light units. However, the user connected to the nearest light unit may experience poor channel performance (i.e., high level of CCI). Thus, for the NI-Rs, when two users are located near to each other, the NI-Rs of the two users view the same light units. This leads to high CCI level at these users. In this case, we choose to give the priority to the stationary receiver (no data will be transmitted to the mobile user). On the other hand, for the NI-ADR, the controller assigns the best light unit to the stationary user and the second best light unit to the mobile user. This is due to the NI-ADR construction where there are many faces and each face is directed to a different direction, which enables NI-ADR to see many light units from a different direction.

*3.2 Performance Analysis of the Data Channels*

To find the data rate that can be provided from each active light unit (the light units allocated to users), we consider that each light unit sends data with BER = $10^{-6}$, which gives a reliable connection between transmitters and receivers. The relationship between the $BER$ and signal to interference to noise ($SINR$) ratio of the OOK modulation is given as [33]:

$$BER = Q(\sqrt{SINR}) \quad (16)$$

where $Q(x) = \frac{\int_x^\infty e^{-z^2/2}\,dz}{\sqrt{2\pi}}$. The $SINR$ can be expressed as [23], [34]:

$$SINR = \frac{R^2 (P_{s1} - P_{s0})^2}{\sigma_{td}^2 + I} \quad (17)$$

where ($P_{s1}$) is the received optical power associated with logic 1, ($P_{s0}$) is the received optical power associated with logic 0, $\sigma_{td}$ is the standard deviation of the total noise associated with the received data and $I$ is the CCI level. To calculate $\sigma_{td}$, we considered the receiver bandwidth while we considered the bandwidth of the BPF to obtain $\sigma_t$. All photodetectors in the NI-R and in each face of the NI-ADR views the same area. Hence, the amount of the CCI level that was calculated from the green channel (SCM tone) is used to calculate the level of the CCI of the signal channels. In addition, each photodetector was covered by a specific optical filter. Thus, the level of the CCI on the data channels can be obtained from the CCI level of the SCM tones as:

$$I_{red} = \left(\frac{R_r}{R_g}\right)^2 \left(\frac{Pt_r}{Pt_g}\right) I_{green},$$
$$I_{yellow} \left(\frac{R_y}{R_g}\right)^2 \left(\frac{Pt_y}{Pt_g}\right) I_{green} \text{ and} \quad (18)$$
$$I_{blue} \left(\frac{R_b}{R_g}\right)^2 \left(\frac{Pt_b}{Pt_g}\right) I_{green}$$

where $R_r$, $R_y$ and $R_b$ are responsivity of the photodetector for the red, yellow and blue colours and $Pt_r$, $Pt_g$, $Pt_y$ and $Pt_b$ are the red, green, yellow and blue optical transmitted power. Consequently, the $SINR$ of each data channel is written as:

$$SINR_{red} = \frac{R_r^2 (P_{s1} - P_{s0})^2}{\sigma_{td}^2 + \left(\frac{R_r}{R_g}\right)^2 \left(\frac{Pt_r}{Pt_g}\right) I_{green}} \quad (19)$$

$$SINR_{yellow} = \frac{R_y^2 (P_{s1} - P_{s0})^2}{\sigma_{td}^2 + \left(\frac{R_y}{R_g}\right)^2 \left(\frac{Pt_y}{Pt_g}\right) I_{green}} \quad (20)$$

$$SINR_{blue} = \frac{R_b^2 (P_{s1} - P_{s0})^2}{\sigma_{td}^2 + \left(\frac{R_b}{R_g}\right)^2 \left(\frac{Pt_b}{Pt_g}\right) I_{green}} \quad (21)$$

$$SINR_{green} = \frac{R_g^2 (P_{s1} - P_{s0})^2}{\sigma_{td}^2 + I_{green}} \quad (22)$$

It should be noted that for the data signals, we considered the level of the CCI that comes from the active light units only as other units do not transmit data.

**4. Simulation results**

Three users (two stationary users and one mobile user) were investigated in four scenarios as shown in Fig. 5. In the first scenario, one of the stationary users (user 1) was located at (1 m, 1 m, 1 m) while the other stationary user (user 2) was placed at (1 m, 7 m, 1 m). In the second scenario, user 1 was placed at (1 m, 4 m, 1 m) and user 2 was located at (3 m, 4 m, 1 m). In the third scenario, user 1 and user 2 were located at (2 m, 1 m, 1 m) and (2 m, 7 m, 1 m), respectively while for the fourth scenario, user 1 was placed at (1 m, 1 m, 1 m) and user 2 was placed at (2 m, 4 m, 1 m). The mobile user (user 3) moved in steps of 1 m along the *y*-axis and at *x* = 0.5 m and $x = 1.5$ m in all scenarios. The four scenarios were selected out of many cases based on many criteria such as separation distance between the optical receivers, the effect of the mobility on the performance and the weakest points on the communication floor (the largest distances between the transmitters and the receivers).

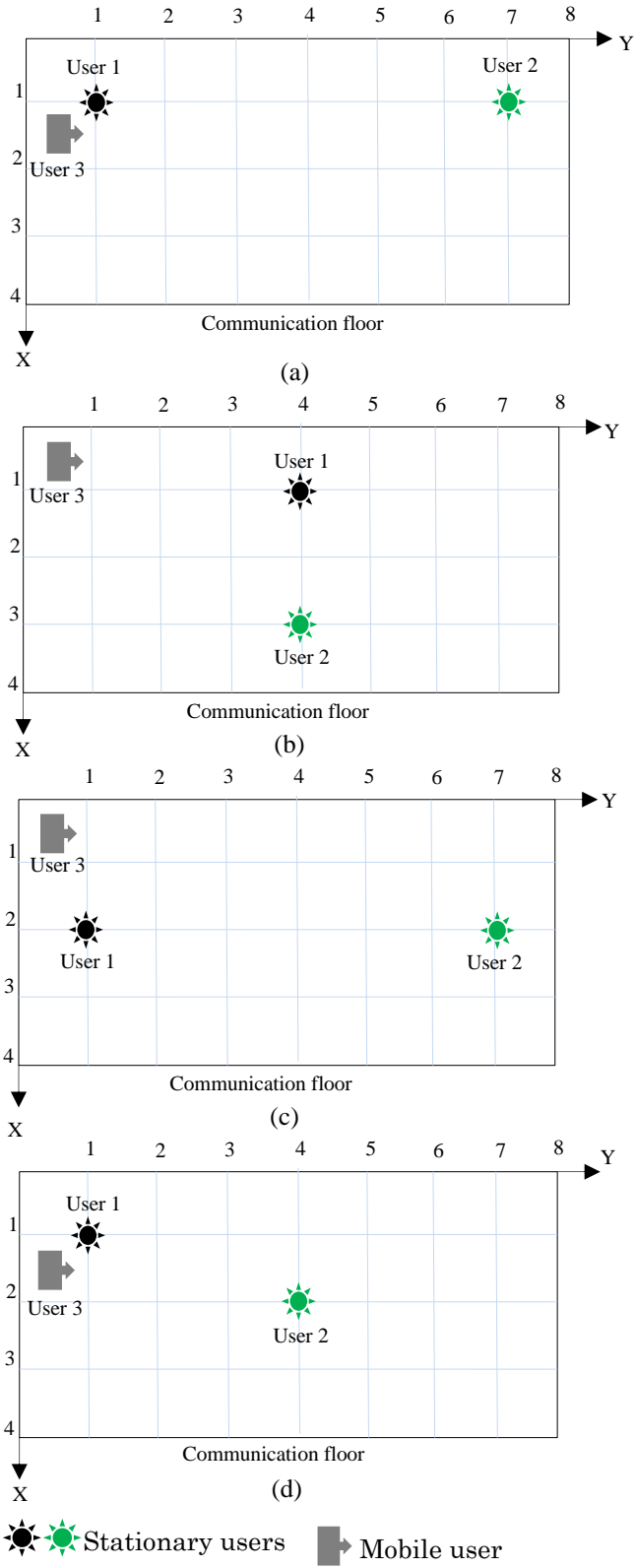

Fig. 5: Receivers location on the communication floor: (a) scenario 1, (b) scenario 2, (c) scenario 3 and (d) scenario 4.

Fig. 6 and Fig. 7 show the aggregate data rate of the NI-R and the NI-ADR for the three users under the four different scenarios. The results were obtained when the mobile user (user 3) moves along the *y*-axis and at $x = 0.5$ m and $x = 1.5$ m. It should be noted that the maximum data rate of each data channel from each assigned light unit for each user was obtained when the BER = $10^{-6}$ (SINR = 13.6 dB). The maximum achieved data rate of each data channel depends on many factors. These are 1) the distance between the user and its allocated light unit, 2) the level of CCI at the user and 3) the diffuse reflections, which introduce ISI. When two users are located at the same location, the priority is given to the stationary user (the controller assigns the best light unit to the stationary user). Thus, for the NI-R, when the mobile user is near to one of the stationary users, the achieved data rate of the mobile user was almost equal to zero at BER = $10^{-6}$ as shown in Fig. 6 and Fig. 7. This is due to the high level of CCI compared to its signal channel. In this case, the controller assigns light units to just the stationary users. On the other hand, by using the NI-ADR, better performance was achieved as the NI-ADR has many faces and each face is directed to a different area, which enables it to view many light units at its location. Thus, two users can be served when they are located at the same place by using the NI-ADR as shown in Fig. 6 and Fig. 7. In addition, the achieved data rate of the NI-ADR was better than the data rate of the NI-R. This is attributed to the small FOV (20º) of each photodetector at each branch of the NI-ADR, which means that the number of rays due to first and second reflections that were captured by each photodetector in each branch is small compared with that of the NI-R. This leads to a reduction in the level of the CCI and consequently increase in the data rate. It should be noted that the 3-dB channel bandwidth of each system using a given colour (see Table 2) is higher than the obtained data rate of each channel for NI-R and NI-ADR. The full channel bandwidth could not be exploited due to the limited transmit power available. High power per colour may be used to increase the data rate, but this affects the illumination. In addition, our system can support up to 12 users (as we installed 12 light units in the ceiling of the room) in the room if each user is given all four colours in the light unit. However, the system can support up to 36 users in the room if each user is given one colour from its optimum light unit. In this case the, it is helpful if the users are distributed uniformly in the room with appropriate distances between them. Adding more users to the systems increases the aggregate data rate of the system. However, this degrades the per-user data rate due to interference.

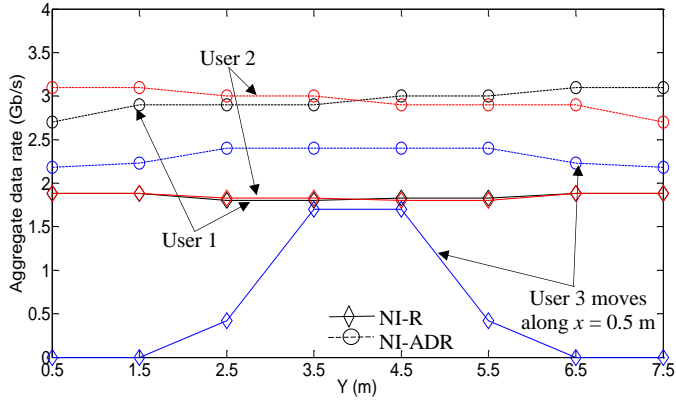
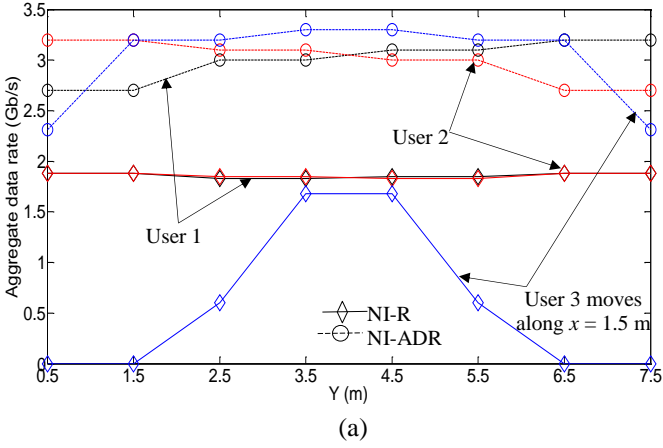

(a)

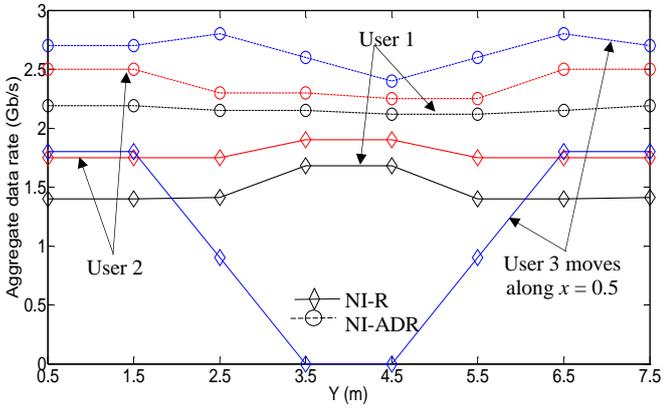
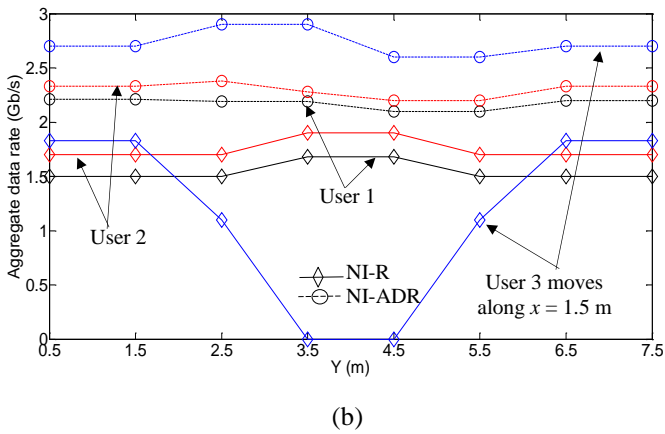

(b)

Fig. 6: Aggregate data rate of the NI-R and the NI-ADR for the three users when the mobile user moves along $x = 0.5$ m and $x = 1.5$ m along the y-axis and under: (a) scenario 1 and (b) scenario 2.

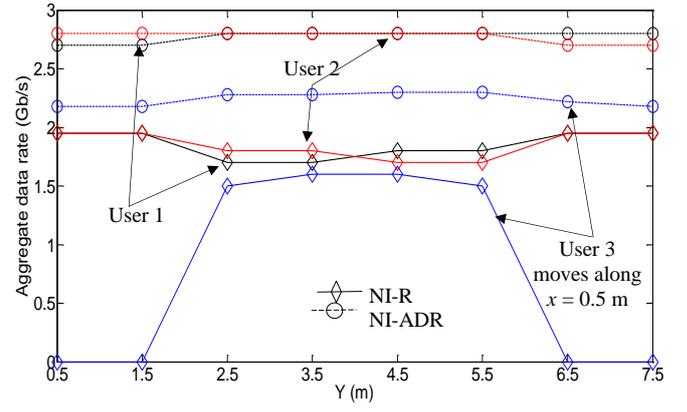
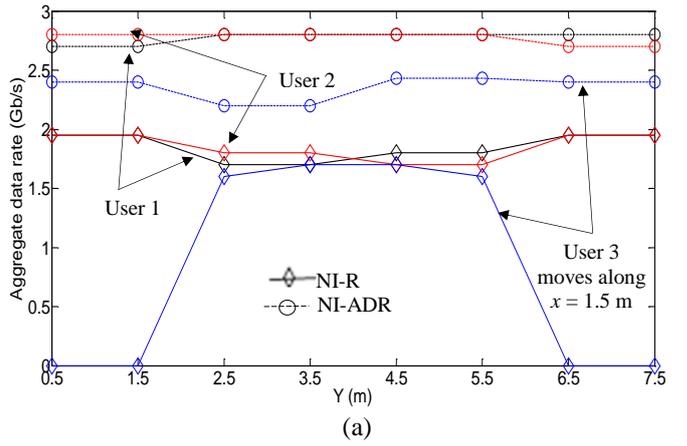

(a)

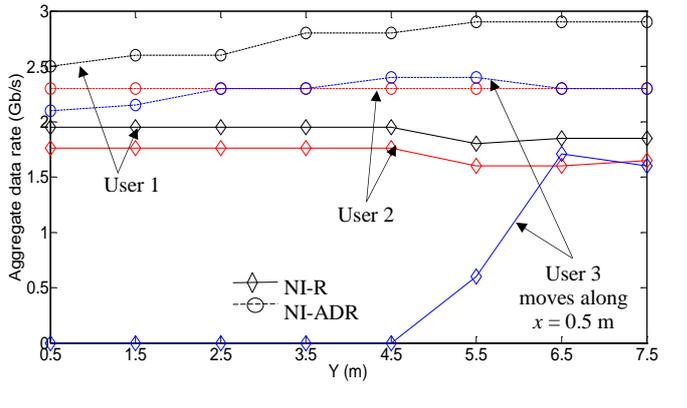
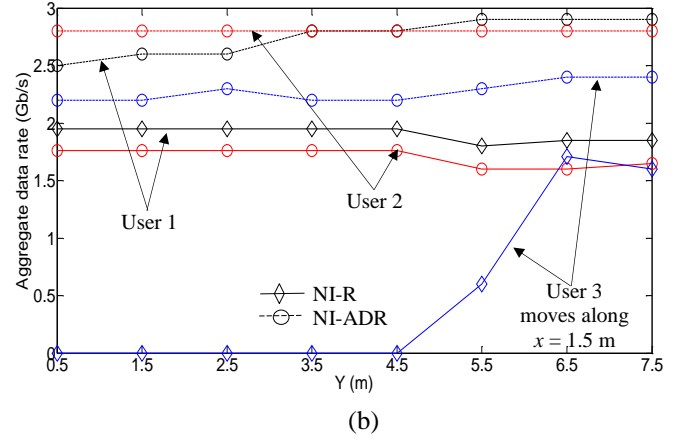

(b)

Fig. 7: Aggregate data rate of the NI-R and the NI-ADR for the three users when the mobile user moves along $x = 0.5$ m and $x = 1.5$ m along the y-axis and under: (a) scenario 3 and (b) scenario 4.

## 5. Conclusions

A multi-user VLC system based on SCM tones and WDM was introduced in this paper. The SCM tones were used to give an ID to each light unit, to find the optimum light unit for each user and to calculate the level of the CCI between the light units. The performance of our proposed system was evaluated in an empty room with three users (two stationary users and one mobile user) under four different scenarios. We investigated the performance of the proposed system with two types of receivers (NI-R and NI-ADR) in the presence of CCI, diffuse reflections and mobility. The results showed that when using the NI-R, the distances between users should be considered to avoid high CCI between the users. However, the NI-ADR eliminates this limitation in the NI-R. In addition, a higher data rate was obtained when the NI-ADR was used compared with the NI-R. This is attributed to the small FOV of the NI-ADR, which limits the number of captured rays due to first and second reflections and consequently reduces the effect of the CCI and diffuse reflections induced ISI.